\documentclass[aps,prl,showpacs,reprint,groupedaddress]{revtex4-1}

\usepackage{amsmath}
\usepackage{amssymb,amsthm}
\usepackage{verbatim}
\usepackage{epsfig}

\begin{document}


\title{
Constructing all entanglement witnesses from density matrices}


\author{Bang-Hai Wang$^{1,2}$ and Dong-Yang Long$^1$}%
\affiliation{%
$^1$Department of Computer Science, Sun Yat-sen University, Guangzhou 510006, People's Republic of China
}%
\affiliation{%
$^2$Faculty of Computer, Guangdong University of Technology, Guangzhou 510006, People's Republic of China
}%


\date{\today}

\begin{abstract}
We demonstrate a general procedure to construct entanglement witnesses for any entangled state. This procedure is based on the trace inequality and a general form of entanglement witnesses, which is in the form $W=\rho-c_{\rho} I$,
where $\rho$ is a density matrix, $c_{\rho}$ is a non-negative number related to $\rho$, and $I$ is the identity
matrix. The general form of entanglement witnesses
is deduced from Choi-Jamio{\l}kowski isomorphism, that can be
reinterpreted as that all quantum states can be obtained by a maximally
quantum entangled state pass through certain completely positive
maps. Furthermore, we provide the necessary and sufficient condition of
the entanglement witness $W=\rho-c_{\rho}I$ in operation, as well as
in theory.
\end{abstract}


\maketitle

Quantum entanglement, which is applied to various types of quantum
information processing such as quantum computation\cite{Shor94},
quantum dense coding\cite{Bennett92}, quantum
teleportation\cite{Bennett93}, quantum cryptography\cite{Ekert91},
etc., has been incorporated as a central notion in quantum
information theory\cite{Horodecki09}. It is well known that
entanglement can be identified by applying all positive but not
completely positive (PNCP) maps to a given
state\cite{Horodecki96,Sperling09}. However, it is not easy to find
and  physically realize PNCP maps\cite{Yu05}. An equivalent approach
of identifying entanglement is based on entanglement witnesses
(EWs)\cite{Horodecki09,Sperling09}. EWs are observables that completely characterize separable (not
entangled) states and allow us to detect entanglement
physically\cite{Horodecki09}. This make EWs one of the main methods of physically detecting entanglement.

Constructing the EW for an entangled state is a difficult task, and the determination of EWs for all entangled states is a nondeterministic polynomial-time (NP) hard problem\cite{Gurvits04,Doherty04,Hou10a}. Concerning this topic, much work has been done for constructing special EWs (for example, Refs.\cite{Doherty04,Sperling09,Hou10a,Chruscinski09,Chruscinski10}). In this Brief Report, we show a general form
of EWs from density matrices. This general form of entanglement
witnesses is deduced from a known relation; that is, any quantum state can actually be generated
from a maximally entangled quantum state with a completely positive
map. The trace inequality indicates an EW can be built up as commuting with a given entangled state. Therefore, we provide a general procedure for detecting entangled quantum
states. Furthermore, we provide the necessary and sufficient
condition of the entanglement witness $W=\rho-c_{\rho}I$ both in
operation and in theory.

For our purpose, we first consider quantum states on the finite
dimensional Hilbert space
$\mathcal{H}_{AB}=\mathcal{H}_A\otimes\mathcal{H}_B$. Let
$\text{dim}(\mathcal{H}_{A})={d}_{A}$, $\text{dim}(\mathcal{H}_{B})={d}_{B}$ and
$\text{dim}(\mathcal{H}_{AB})={d}_{AB}$. Denote $P_+$ as the density matrix
of the maximally quantum entangled state
$|\beta\rangle=d_{A}^{-1/2}\sum_{i}|i\rangle\otimes|i\rangle$ on
$\mathcal{H}_A\otimes\mathcal{H}_A$, where
$\{|i\rangle\}_{i=0}^{d_A-1}$ are computational bases in
$\mathcal{H}_{A}$, i.e. $P_+=|\beta\rangle\langle\beta|$.

\textit{The relation between any quantum state and a maximally entangled state.} Quantum entanglement, which is a fascinating feature of quantum
theory, underlines the intrinsic order of statistical relations
between subsystems of a compound quantum system\cite{Schrodinger35}.
In the following, we show a relation between all quantum states and
a maximally entangled state. This relation
comes from a well-known feature called ``channel-state duality" or
``Jamio{\l}kowski isomorphism" or ``Choi-Jamio{\l}kowski isomorphism". This result first appeared in Ref.
\cite{Choi75} with a proof. We now show it in a different manner.

\textbf{Lemma 1.} Any matrix $H$ on $\mathcal{H}_{AB}$ is a
Hermitian matrix if and only if it can always be written as
\begin{equation}
H=(I\otimes\Phi)(P_+)
\end{equation}
where $\Phi$ is a hermiticity-preserving linear map.

\textbf{Lemma 2.} (\cite{filterhor}) A linear map $\Lambda$:
$\mathcal{H}{_{A}}\rightarrow \mathcal{H}{_{B}}$ is completely
positive if and only if the matrix $D\in \mathcal{H}_{AB}$ given by
\begin{equation}
D=(I\otimes \Lambda)(P_+)\label{oper}
\end{equation}
is positive semidefinite.

By Lemmas 1 and 2, we can get the following result.

\textbf{Theorem 1.} Any matrix $\rho$ on $\mathcal{H}_{AB}$ is a
bipartite density matrix if and only if it can always be written as
\begin{equation}
\rho=(I \otimes\Lambda)(P_+), \label{rhocp}
\end{equation}
where $\Lambda\colon\mathcal{H}_A\rightarrow\mathcal{H}_B$ is a
completely positive map.

Note that $\Lambda$ may not be trace preserving and $\rho$ may not
be normalized. A similar result on
$\mathcal{H}_A\otimes\mathcal{H}_A$ was shown by DiVincenzo \emph{et
al}. \cite{DiVincenzo00}. If $\Lambda$ is trace preserving, $\Lambda$
is an entanglement-breaking channel (EBC)\cite{M.Horodecki03}, but not
vice versa \cite{Wang11}.

\textit{The general form of entanglement witnesses.} Starting from the positive map\cite{Horodecki96}, the concept of
EW was applied to detecting the presence of
entanglement\cite{Terhal00}. EWs are observables
whose expectation value can reveal something about the entanglement
in a given state \cite{Vedral08}.  A Hermitian matrix $W=W^\dagger$
on $\mathcal{H}_{AB}$ is an EW if it has (i)
at least one negative eigenvalue and (ii) nonnegative mean values in
all separable quantum states, or equivalently satisfy
\begin{equation}
\langle\mu_A\nu_B|W|\mu_A\nu_B\rangle\geq 0
\end{equation} for all pure product states
$|\mu_A\nu_B\rangle$\cite{Horodecki09,Lewenstein00,Yu05}. If
we have negative mean value in a quantum state for an EW, the quantum state is entangled. In that case, we say that
the EW ``witnesses" (detects) the quantum state. To
balance out the ``not trace-preserving" property  of the completely
positive map in this Brief Report, we need another property of
EWs: (iii) if $W$ is an EW, $\gamma W$ keeps all
properties of $W$ as an EW for a non-negative
number $\gamma$. Note that the third property is different from the
definition by Lewenstein \emph{et al}. for comparing the action of
different EWs\cite{Lewenstein00}.

\textbf{Theorem 2.} Any bipartite density matrix $\pi$ is entangled
if and only if there exists a density matrix $\rho$ and a
non-negative number $c_{\rho}$ such that the matrix
\begin{equation}
W=\rho-c_{\rho} I\label{ew-density0}
\end{equation}
satisfies $\text{tr}(W\pi)<0$ and $\text{tr}(W\sigma)\geq 0$ for all
separable states $\sigma$.

This result shows that every entangled state in a composite system
has an EW in the simple form $W=\rho-c_{\rho}I$,
where $c_{\rho}$ is a non-negative number and $\rho$ is a density
matrix. This result also shows that the research on density matrices
can replace the research on EWs since $c_{\rho}I$
is simple.

 \textbf{Proof.} By Lemma 1, any EW $W'$ can be written as
\begin{equation}
W'=(I\otimes\Theta)(P_+),
\end{equation}
where $\Theta$ is a positive map\cite{Terhal00,Korbicz08}. By
property (iii) of EWs,
\begin{equation}
W=(1-p)W'=(1-p)(I\otimes\Theta)(P_+)\label{sameEW}
\end{equation} is the same EW as $W'$ for
$0<p<1$.

We could mix $(1-p)\Theta$ with a simple completely
positive map $p\Lambda_s$: $(1-p)\Theta+p\Lambda_s$.
By structural completely positive approximation (SCPA) and
structural physical approximation (SPA)\cite{Horodecki02,Korbicz08}
for proper $p$,
\begin{equation}
 [I\otimes ((1-p)\Theta+p\Lambda_s)](P_+)=I\otimes\Lambda'(P_+)=\rho,\label{ew-density}
\end{equation}where $\Lambda'=(1-p)\Theta+p\Lambda_s$ is a completely positive map and $\rho$ is a density matrix by Theorem
1. Note that $\Lambda'$ could be not trace preserving. Rewriting Eq.
(\ref{ew-density}),
\begin{equation}
\rho-p(I\otimes\Lambda_s)(P_+)=(1-p)(I\otimes
\Theta)(P_+)=(1-p)W'.\label{ewmap}
\end{equation}

Without loss of generality, let $\Lambda_s(\cdot)=\sum_{ij} E_{ij}
(\cdot)E_{ij}^\dagger$, where $E_{ij}=|i\rangle\langle j|$ and
$\{|i\rangle\}_{i=0}^{d_A-1}$ are computational bases in
$\mathcal{H}_{A}$, $\{|j\rangle\}_{i=0}^{d_B-1}$ in
$\mathcal{H}_{B}$\cite{Korbicz08}. We have $I\otimes
p\Lambda_s(P_+)=\frac{pI}{d_{AB}}$. By Eqs. (\ref{sameEW}) and (\ref{ewmap}), we
have
\begin{equation}
W=\rho-p(I\otimes\Lambda_s)(P_+) = \rho- c_{\rho} I.
\end{equation} where $c_{\rho}$ is a
non-negative number.
$\blacksquare$\hfill

Let $F=\rho+(1-c_{\rho})I$. By Eq. (\ref{ew-density0}), $W=I-F$. This is
the form of EWs in Refs. \cite{Sperling09} and \cite{Hou10}.
Clearly, all EWs in \cite{Hou10} can be constructed in the form of
$W=\rho-c_{\rho}I$. A general discussion can be found in Refs.
\cite{Sperling09} and \cite{Hou10}. Moreover, we can also easily obtain Theorem 2 mathematically from the EW in the form $P-cI$ (Hermitian matrix), where $P$ is a positive matrix. However, they all have no physical interpretation.

We can prove that $\lambda^{\text{min}}(\rho)<c_{\rho}\leq d^{\text{min}}(\rho)$
if $W=\rho-c_{\rho} I$ is an EW, where $\lambda^{\text{min}}(\rho)$ and
$d^{\text{min}}(\rho)$ are the minimum eigenvalue of $\rho$ and the minimum
diagonal element in $\rho$, respectively. However, for any density
matrix $\rho$, such as the diagonal state (its density matrix is a diagonal matrix), the EW in the form
$W=\rho-c_{\rho}I$ does not always exist. Generally, it is not easy
for any density matrix $\rho$ to find $c_{\rho}$ to make
$W=\rho-c_{\rho} I$ an EW, but we have the following
operational result.

\textbf{Theorem 3.} A Hermitian matrix $W=\rho-c_{\rho}I$ is an EW
for any density matrix $\rho$, if $\lambda^{\text{min}}(\rho)<c_{\rho}\leq
c_\rho^{\text{max}}=\sum_{ij}|d_i|^2|f_j|^2\rho_{ijij}-\sum_{i,j<l}2(|d_i|^4+|f_j|^2|f_l|^2)|Re(\rho_{ijil})|
-\sum_{i<k,jl}2(|d_i|^2|d_k|^2+|f_j|^2|f_l|^2)|Re(\rho_{ijkl})|$,
where $\sum_i^{d_A}|d_i|^2=1$, $\sum_j^{d_B}|f_j|^2=1$ and
$\text{Re}(\rho_{ijkl})$ is the real part of the element $\rho_{ijkl}$ of
$\rho$.

Since any density matrix $\rho$ can be written as
$\rho=\sum_{ijkl}a_{ijkl}+\sum_{ijkl,i\neq k\&j\neq l}b_{ijkl}i$,
where $a_{ijkl}$ and $b_{ijkl}$ are real, $b_{ijkl}=-b_{klij}$, and
$tr[(|\mu_A\nu_B\rangle\langle\mu_A\nu_B|)\sum_{ijkl,i\neq
k\&j\neq l}b_{ijkl}]=0$, we can only consider real parts of a
density matrix.

 \textbf{Proof.} Any density matrix $\rho$ on $\mathcal{H}_{AB}$ can be defined as
\begin{eqnarray}
\rho=\sum_{ijkl}\langle ij|\rho|kl\rangle(|i\rangle\langle k|\otimes
|j\rangle\langle l|)\label{whermitian}
\end{eqnarray}
by computational (real orthonormal) bases
$\{|i\rangle\}_{i=0}^{d_A-1}$ and $\{|k\rangle\}_{i=0}^{d_A-1}$ in
$\mathcal{H}_{A}$, and $\{|j\rangle\}_{i=0}^{d_B-1}$ and
$\{|l\rangle\}_{i=0}^{d_B-1}$ in $\mathcal{H}_{B}$.

For any unit product vector
$|\mu_A\nu_B\rangle=\sum_{ij}d_if_j|i\rangle |j\rangle$ on
$\mathcal{H}_{AB}$ with $\sum_i^{d_A}|d_i|^2=1$ and
$\sum_j^{d_B}|f_j|^2=1$, by Eq. (\ref{whermitian}),
\begin{eqnarray}
&&\text{tr}[\rho(|\mu_A\nu_B\rangle\langle\mu_A\nu_B|)]\nonumber\\
&&=\langle\mu_A\nu_B|\rho|\mu_A\nu_B\rangle\\
&&=\langle\mu_A\nu_B|[\sum_{ijkl}\langle ij|\rho
|kl\rangle(|i\rangle\langle k|\otimes |j\rangle\langle
l|)]|\mu_A\nu_B\rangle\\
&&=\sum_{ijkl}d_i^*f_j^*d_kf_l\rho_{ijkl}\\
&&=\sum_{i<k,jl}2Re(d_i^*f_j^*d_kf_l)\text{Re}(\rho_{ijkl})\nonumber\\
&&+\sum_{i=k,j<l}2Re(d_i^*f_j^*d_kf_l)\text{Re}(\rho_{ijkl})\nonumber\\
&&+\sum_{i=k,j=l}Re(d_i^*f_j^*d_kf_l)\text{Re}(\rho_{ijkl})\\
&&=\sum_{i<k,jl}2[\text{Re}(d_i^*f_j^*d_kf_l)+|d_i|^2|d_k|^2+|f_j|^2|f_l|^2-|d_i|^2|d_k|^2\nonumber\\
&&-|f_j|^2|f_l|^2]\text{Re}(\rho_{ijkl})+\sum_{i,j<l}2\text{Re}(d_i^*f_j^*d_if_l)\text{Re}(\rho_{ijil})\nonumber\\
&&+\sum_{ij}|d_i|^2|f_j|^2\rho_{ijij}\\
&&\geq
\sum_{ij}|d_i|^2|f_j|^2\rho_{ijij}-\sum_{i,j<l}2(|d_i|^4+|f_j|^2|f_l|^2)|\text{Re}(\rho_{ijil})|\nonumber\\
&&-\sum_{i<k,jl}2(|d_i|^2|d_k|^2+|f_j|^2|f_l|^2)|\text{Re}(\rho_{ijkl})|,
\end{eqnarray}
where $\rho_{ijij}$ are the diagonal elements of $\rho$. Thus, if
$W=\rho-c_{\rho}I$, $\text{tr}(W\sigma)\geq 0$ for all separable
states $\sigma$.$\blacksquare$\hfill

\textit{A general procedure of detecting entangled states.} Generally, for any entangled
state $\pi$, it is not easy to find $\rho$ to make
$W=\rho-c_{\rho}I$ detecting it, but we
have the following operational result. Let us recall a well-known
trace inequality for Hermitian matrices.

\textbf{Lemma 3.} (\cite{Marshall79,Horn90}) For any two Hermitian matrices $H,K$ in $\mathcal{H}_{AB}$,
\begin{eqnarray}
\sum_{i=0}^{d_{AB}-1}\lambda_i(H)\lambda_{d_{AB}-i}(K)\leq
tr(HK) \leq\sum_{i=0}^{d_{AB}-1}\lambda_i(H)\lambda_i(K),\nonumber
\end{eqnarray} where $\lambda_1(H)\geq\cdots\geq\lambda_{d_{AB}-1}(H)$, $\lambda_1(K)\geq\cdots\geq\lambda_{d_{AB}-1}(K)$, and $\lambda_i(H)$ and $\lambda_i(K)$ are the
eigenvalues of $H$ and $K$, respectively.

If $W=\rho-c_{\rho}I$ is the EW for any entangled density matrix
$\pi$, $\text{tr}(W\pi)=\text{tr}(\rho\pi)-c_{\rho}<0$. It requires $\text{tr}(\rho\pi)$
as small as possible and $c_{\rho}$ as big as possible to make
$\text{tr}(\rho\pi)-c_{\rho}<0$. By Lemma 3, it is not difficult to
conclude that the minimum of $\text{tr}(\rho\pi)$ is equal to
$\sum_{i=0}^{d_{AB}-1}\lambda_i(\rho)\lambda_{d_{AB}-i}(\pi)$ if and
only if $\rho$ and $\pi$ are simultaneously
diagonalizable\cite{Marshall79,Horn90,Nielsen00}. Therefore, we have the following result.

\textbf{Theorem 4.} An EW can be built up as commuting with a given entangled state.

By Theorem 4, we have a
general procedure of constructing EW for any density matrix.

(i) Suppose the spectral decomposition of any density matrix $\pi=\sum_i
\lambda_i|\psi_i\rangle\langle\psi_i|$ with $\lambda_0\leq
\lambda_1\leq \cdots\leq \lambda_{d_{AB}-1}$. (ii) Construct
$\rho=\sum_i \gamma_{d_{AB}-1-i}|\psi_i\rangle\langle\psi_i|$ with
$0\leq\gamma_0\leq \gamma_1\leq \cdots\leq
\gamma_{d_{AB}-1}$ different from $\lambda_0\leq
\lambda_1\leq \cdots\leq \lambda_{d_{AB}-1}$. (iii) Compute $c_\rho^{\text{max}}$ by Theorem 3. (iv) Let $W=\rho-c_\rho^{\text{max}} I$ be the EW. (v) If
$tr(W\pi)<0$, $\pi$ is entangled, otherwise repeat (ii).

Let us consider a simple example. Suppose
$\pi_p=p|\psi\rangle\langle\psi|+(1-p)I/4$, where
$|\psi\rangle=\frac{1}{\sqrt{2}}(|00\rangle+|11\rangle)$ and
$0<p<1$. It is well known that $\pi_p$ is an entangled state if
$p>\frac{1}{3}$ with positive partial transposition (PPT) criterion\cite{Horodecki96}. Let us detect
$\pi_p$ with the general procedure above. (i) The spectral decomposition
$\pi_p=\frac{1+3p}{4}|\omega_1\rangle\langle\omega_1|+\frac{1-p}{4}|\omega_2\rangle\langle\omega_2|+\frac{1-p}{4}|\omega_3\rangle\langle\omega_3|+\frac{1-p}{4}|\omega_4\rangle\langle\omega_4|$,
where $|\omega_1\rangle=\frac{1}{\sqrt{2}}\{1,0,0,-1\},
|\omega_2\rangle=\frac{1}{\sqrt{2}}\{1,0,0,1\},
|\omega_3\rangle=\{0,1,0,0\}, |\omega_4\rangle=\{0,0,1,0\}$.
(ii) Construct
$\rho_q=\frac{1+3q}{4}|\omega_1\rangle\langle\omega_1|+\frac{1-q}{4}|\omega_2\rangle\langle\omega_2|+\frac{1-q}{4}|\omega_3\rangle\langle\omega_3|+\frac{1-q}{4}|\omega_4\rangle\langle\omega_4|$
and $-\frac{1}{3}\leq q<0$. It is interesting that
$\{\frac{1+3p}{4},\frac{1-p}{4},\frac{1-p}{4},\frac{1-p}{4}\}\succ
\{\frac{1-q}{4},\frac{1-q}{4},\frac{1-q}{4},\frac{1+3q}{4}\}$ for
$p>\frac{1}{3}$ and $-\frac{1}{3}\leq q<0$, where ``$x\succ y$" means
$x$ majorizes $y$ \cite{Marshall79,Nielsen00,Horn90}. (iii) We can compute
$c_{\rho_q}^{\text{max}}=\frac{1+q}{4}$ by Theorem 3 (see Appendix). (iv) Let $W=\rho_q-\frac{1+q}{4}I$. (v) $\text{tr}(W\rho_p)=\frac{(3p-1)q}{4}<0$ if and only if $p>\frac{1}{3}$
and $-\frac{1}{3}\leq q<0$. Therefore, $\pi_p$ is entangled if and only if
$p>\frac{1}{3}$. Let $c_{\rho_q}=\frac{1+2q}{4}$. Interestingly, $W=\rho_q-\frac{1+2q}{4}I$ can detect $\pi_p$ for $p>\frac{2}{3}$.

Note that we need to construct eigenvectors besides eigenvalues of
$\rho$ if there exists one or more  zero eigenvalues of $\pi$ ,
because only if the multiplicity of each eigenvalue of $\pi$ is one,
the spectral decomposition of $\pi$ will be unique.
For example, Bell state $|\psi\rangle=\frac{1}{\sqrt2}(|00\rangle+|11\rangle)$. Its density matrix $P_+^B=|\psi\rangle\langle\psi|$. Construct $\rho=a|\psi\rangle\langle\psi|+b|01\rangle\langle 01|+b|10\rangle\langle 10|+b|\phi\rangle\langle\phi|$, where $|\phi\rangle=\frac{1}{\sqrt2}(-|00\rangle+|11\rangle)$, $b>a>0$ and $a+3b=1$. We can compute $W=\rho-\frac{a+b}{2}I$ and $\text{tr}(WP_+^B)<0$.

Note that
$c_{\rho}$ may not exist for the density matrix $\rho$ and
$W=\rho-c_{\rho} I$ may not be the entanglement witness that can
detect the given $\pi$. The procedure never stops if the given state is not
entangled (separable). By considering a countable set of product vectors spanning the range of a given state, Hulpke and Bru{\ss}\cite{Hulpke05} construct an algorithm (procedure) to detect the separability of the state, in which its termination is guaranteed if the state is separable. Similar to \cite{Hulpke05}, we can run both the procedure
above and the Hulpke and Bru{\ss}\cite{Hulpke05} algorithm for the detection of a separable state in
parallel with detecting entanglement. A general discussion can be found in Refs.
\cite{Horodecki09,Doherty04,Hulpke05}.

It should be stressed that there is no universal EW that
detects all entangled states, and there is no general procedure for
constructing EWs \cite{Chruscinski09}. Many problems on the procedure above are under investigation.

\textit{The necessary and sufficient condition in theory.} What is the necessary and sufficient condition for
$W=\rho-c_{\rho}I$ to be an EW in theory? Let us start from the
following results.

\textbf{Theorem 5.} For any density matrix $\rho$ with spectral
decomposition $\rho=\sum_r \lambda_r|\psi_r\rangle\langle\psi_r|$ if
$c_{\rho}\leq c_\rho^{\text{max}}$ for any unit product vector
$|\mu_A\nu_B\rangle$, $\text{tr}(W\sigma)\geq 0$ for all
separable states $\sigma$, where
$c_\rho^{\text{max}}=inf_{\parallel\mu_A\parallel=1,
\parallel\nu_B\parallel=1}\sum_r
\lambda_r\parallel\langle\psi_r|\mu_A\nu_B\rangle\parallel^2$
and $W=\rho-c_{\rho}I$.

 \textbf{Proof.} For any unit product vector
$|\mu_A\nu_B\rangle$, by Eq. (\ref{ew-density0}) and
$\rho=\sum_r \lambda_r|\psi_r\rangle\langle\psi_r|$, we have
\begin{eqnarray}
&&\text{tr}[W(|\mu_A\nu_B\rangle\langle\mu_A\nu_B|)]\nonumber\\
&=&\langle\mu_A\nu_B|\rho|\mu_A\nu_B\rangle-c_{\rho}\\
&=&\langle\mu_A\nu_B|(\sum_r \lambda_r|\psi_r\rangle\langle\psi_r|)|\mu_A\nu_B\rangle-c_{\rho}\\
&=&\sum_r
\lambda_r\parallel\langle\psi_r|\mu_A\nu_B\rangle\parallel^2-c_{\rho}\\
&\geq&c_\rho^{\text{max}}-c_{\rho}\geq 0,
\end{eqnarray} where $c_\rho^{\text{max}}=inf_{\parallel\mu_A\parallel=1,
\parallel\nu_B\parallel=1}\sum_r
\lambda_r\parallel\langle\psi_r|\mu_A\nu_B\rangle\parallel^2$.
Thus, if $W=\rho-c_{\rho}I$, $\text{tr}(W\sigma)\geq 0$ for all
separable states $\sigma$.$\blacksquare$\hfill

\textbf{Corollary 1.} For any density matrix $\rho$,
$W=\rho-c_{\rho}I$ is an EW if and only if
$\lambda^{\text{min}}(\rho)<c_{\rho}\leq c_\rho^{\text{max}}$, where
$\lambda^{\text{min}}(\rho)$ is the the minimum eigenvalue and
$c_\rho^{\text{max}}=inf_{\parallel\mu_A\parallel=1,
\parallel\nu_B\parallel=1}\sum_r
\lambda_r\parallel\langle\psi_r|\mu_A\nu_B\rangle\parallel^2$.


Before giving our result in theory, we need the following lemmas.

\textbf{Lemma 4.} (\cite{Hou98}) A linear map
$\Theta\colon\mathcal{H}_A\rightarrow\mathcal{H}_B$ is positive if
and only if there exists $C_{0}$,...,$C_{k-1},D_{0}$,...,$D_{l-1}$ on $\mathcal{H}_{AB}$ such that $\{D_j\}_{j=0}^{l-1}$ are
a contractive locally linear combination of $\{C_i\}_{i=0}^{k-1}$
and
\begin{equation}
\Theta(X)=\sum_{i=0}^{k-1}C_iXC_i^\dagger-\sum_{j=0}^{l-1}D_jXD_j^\dagger\label{positivemap}
\end{equation}for all $X$ on $\mathcal{H}_B$.

\textbf{Lemma 5.} {\cite{Hou98}} Furthermore, $\Theta$ in Eq.
(\ref{positivemap}) is completely positive if and only if
$\{D_j\}_{j=0}^{l-1}$ is a linear combination of
$\{C_i\}_{i=0}^{k-1}$ with a contractive coefficient matrix.

\textbf{Theorem 6.} A Hermitian matrix $W=\rho-c_{\rho}I$ is an EW
for any density matrix $\rho=(I \otimes\Lambda) (P_+)$ if and only
if $\{\sqrt{d_{AB}c_{\rho}} E_{ij}^{(t)}\}_{t=0}^{l-1}$ is a
contractive locally linear combination of $\{U_r\}_{r=0}^{k-1}$ but
not a linear combination of $\{U_r\}_{r=0}^{k-1}$ with a contractive
coefficient matrix, where $\Lambda (\cdot)=\sum_{r=0}^{k-1} U_r
(\cdot) U_r ^\dagger$, $E_{ij}=|i\rangle\langle j|$ and
$\{|i\rangle\}_{i=0}^{d_A-1}$ are computational bases in
$\mathcal{H}_{A}$, $\{|j\rangle\}_{i=0}^{d_B-1}$ in
$\mathcal{H}_{B}$.

\textbf{Proof.} Any density matrix can be written as
\begin{equation}
\rho=\sum_r \lambda_r|\psi_r\rangle\langle\psi_r|
\end{equation} by means of its spectral decomposition with
nonnegative eigenvalues $\lambda_r$. Taking $V_r$ such that
$I\otimes V_r|\beta\rangle=|\psi_r\rangle$, any density matrix
$\rho$ can be written as
\begin{equation}
\rho=(I \otimes\Lambda)(P_+), \label{rhocp}
\end{equation}
where $\Lambda(\cdot)=\sum_r U_r(\cdot)U_r^\dagger$ is a completely
positive map and $U_r=\sqrt{\lambda_r}V_r$, $I\otimes
V_r|\beta\rangle=|\psi_r\rangle$, $\langle
n|V_r|m\rangle=\sqrt{d_{AB}}a_{mn}^{(r)}$ and
$|\psi_r\rangle=\sum_{mn}a_{mn}^{(r)}|m\rangle|n\rangle$\cite{filterhor}.

By Eq. (\ref{ew-density0}), $W=I \otimes\Lambda(P_+)-I
\otimes\Lambda_s'(P_+)$, where $\Lambda_s'(\cdot)=\sum_t
\sqrt{d_{AB}c_{\rho}}E_{ij}^{(t)}
(\cdot)(\sqrt{d_{AB}c_{\rho}}E_{ij}^{(t)})^\dagger$. By Lemmas 4 and 5, $\{\sqrt{d_{AB}c_{\rho}} E_{ij}^{(t)}\}_{t=0}^{l-1}$ is a
contractive locally linear combination of $\{U_r\}_{r=0}^{k-1} $ but
not a linear combination of $\{U_r\}$ with a contractive coefficient
matrix if and only if $\Phi=\Lambda-\Lambda_s'$ is a PNCP map, and
$W=\rho-c_{\rho} I$ is an EW. $\blacksquare$\hfill

In conclusion, we have demonstrated that any EW can be constructed
from a certain density matrix. This result shows that
the research on density matrices can replace the research on
entanglement witnesses. The trace inequality reveals the
general procedure of constructing EW for any density matrix. Both in
operation and in theory, the necessary and sufficient condition of
an EW in the form $W=\rho-c_{\rho}I$ and some examples are given.
Here we only consider the bipartite case on the finite dimensional
Hilbert space, but we can also generalize our results to
multipartite system and infinite dimensional Hilbert space.

We would like to Thank Professor Guang Ping He, Dan Wu, and D. P. DiVincenzo for
helpful discussions and suggestions. We thank the referee for valuable comments and suggestions to improve the original manuscript. This work is in part
supported by the Key
Project of NSFC-Guangdong Funds ( Grant No. U0935002).


\appendix

\textit{Appendix: The Procedure of Computing $c_{\rho_q}^{\text{max}}$.}

Clearly, \begin{eqnarray} \rho_q=
    \left(
      \begin{array}{cccc}

                 \frac{1+q}{4}  &0 & 0 &\frac{q}{2}  \\

                 0 & \frac{1-q}{4} &0 & 0 \\

                 0  &0 & \frac{1-q}{4} &0 \\

                 \frac{q}{2}&0 &0 &\frac{1+q}{4}
      \end{array}
    \right),
\end{eqnarray}where $-\frac{1}{3}\leq q<0$.

By Theorem 3, $|\mu_A\rangle\nu_B\rangle$ for two qubits can be written as $|\mu_A\rangle\nu_B\rangle=d_0f_0|00\rangle+d_0f_1|01\rangle+d_1f_0|10\rangle+d_1f_1|11\rangle$ with $|d_0|^2+|d_1|^2=1$, $|f_0|^2+|f_1|^2=1$.

\begin{eqnarray}
&&\text{tr}(\rho_q(|\mu_A\rangle\nu_B\rangle\langle\mu_A\langle\nu_B|))=\langle\mu_A\langle\nu_B|\rho_q|\mu_A\rangle\nu_B\rangle\nonumber\\
&=&(d_0^*f_0^*\langle00|+d_0^*f_1^*\langle01|+d_1^*f_0^*\langle10|+d_1^*f_1^*\langle11|)\times\rho_q\times\nonumber\\
&&(d_0f_0|00\rangle+d_0f_1|01\rangle+d_1f_0|10\rangle+d_1f_1|11\rangle)\\
&=&|d_0|^2|f_0|^2\frac{1+q}{4}+|d_0|^2|f_1|^2\frac{1-q}{4}+|d_1|^2|f_0|^2\frac{1-q}{4}\nonumber\\
&&+|d_1|^2|f_1|^2\frac{1+q}{4}+d_0^*f_0^*d_1f_1\frac{q}{2}+d_1^*f_1^*d_0f_0\frac{q}{2}\\
&=&|d_0|^2(|f_0|^2+|f_1|^2)\frac{1+q}{4}+|d_0|^2|f_1|^2\frac{-2q}{4}\nonumber\\
&&+|d_1|^2(|f_0|^2+|f_1|^2)\frac{1+q}{4}+|d_1|^2|f_0|^2\frac{-2q}{4}+2\text{Re}(d_0^*d_1f_0^*f_1)\frac{q}{2}\nonumber\\
&\geq&(|d_0|^2+|d_1|^2)(|f_0|^2+|f_1|^2)\frac{1+q}{4}+[\text{Re}(d_0^*f_1)-\text{Re}(d_1f_0^*)]^2\frac{-q}{2}\nonumber\\
&=&\frac{1+q}{4}+[\text{Re}(d_0^*f_1)-\text{Re}(d_1f_0^*)]^2\frac{-q}{2}\\
&\geq&\frac{1+q}{4}
\end{eqnarray}

Therefore, we have $c_{\rho_q}^{\text{max}}=\frac{1+q}{4}$. In addition, we have another method for computing $c_{\rho_q}^{\text{max}}$.

Any qubit pure state $|\psi\rangle$ can be written as
$|\psi\rangle=\alpha|0\rangle+\beta|1\rangle$, where $\alpha$ and
$\beta$ are complex number and $|\alpha|^2+|\beta|^2=1$. Because
$|\alpha|^2+|\beta|^2=1$, $|\psi\rangle$ can be rewritten as
\begin{eqnarray}
|\psi\rangle=e^{ir}(\cos{\frac{\theta}{2}}|0\rangle+e^{is}\sin{\frac{\theta}{2}}|1\rangle),
\end{eqnarray}
where $\theta$, $r$ and $s$ are real numbers. The factor of $e^{ir}$
out the front can be ignored since it has no observable
effects \cite{Nielsen00}, and for that reason, $|\psi\rangle$ can be
effectively written as
\begin{eqnarray}
|\psi\rangle=\cos{\frac{\theta}{2}}|0\rangle+e^{is}\sin{\frac{\theta}{2}}|1\rangle.
\end{eqnarray}
Therefore, $|\mu_A\rangle\nu_B\rangle$ for two qubits can be written as
\begin{eqnarray}
|\mu_A\rangle\nu_B\rangle&=&(\cos{\frac{\theta_1}{2}}|0\rangle+e^{is_1}\sin{\frac{\theta_1}{2}}|1\rangle)(\cos{\frac{\theta_2}{2}}|0\rangle+e^{is_2}\sin{\frac{\theta_2}{2}}|1\rangle)\nonumber\\
&=&\cos{\frac{\theta_1}{2}}\cos{\frac{\theta_2}{2}}|00\rangle+e^{is_2}\cos{\frac{\theta_1}{2}}\sin{\frac{\theta_2}{2}}|01\rangle\\
&&+e^{is_1}\sin{\frac{\theta_1}{2}}\cos{\frac{\theta_2}{2}}|10\rangle+e^{i(s_1+s_2)}\sin{\frac{\theta_1}{2}}\sin{\frac{\theta_2}{2}}|11\rangle.\nonumber
\end{eqnarray}
\begin{eqnarray}
&&\text{tr}[\rho_q(|\mu_A\rangle\nu_B\rangle\langle\mu_A\langle\nu_B|)]\\
&=&\frac{1+q}{4}\cos^2{\frac{\theta_1}{2}}\cos^2{\frac{\theta_2}{2}}+\frac{q}{2}\cos{\frac{\theta_1}{2}}\cos{\frac{\theta_2}{2}}\sin{\frac{\theta_1}{2}}\sin{\frac{\theta_2}{2}}e^{i(s_1+s_2)}\nonumber\\
&&+\frac{1-q}{4}\cos^2{\frac{\theta_1}{2}}\sin^2{\frac{\theta_2}{2}}+\frac{1-q}{4}\sin^2{\frac{\theta_1}{2}}\cos^2{\frac{\theta_2}{2}}\\
&&+\frac{q}{2}\sin{\frac{\theta_1}{2}}\sin{\frac{\theta_2}{2}}\cos{\frac{\theta_1}{2}}\cos{\frac{\theta_2}{2}}e^{-i(s_1+s_2)}+\frac{1+q}{4}\sin^2{\frac{\theta_1}{2}}\sin^2{\frac{\theta_2}{2}}\nonumber\\
&=&\frac{1+q}{4}\cos^2{\frac{\theta_1}{2}}(\cos^2{\frac{\theta_2}{2}}+\sin^2{\frac{\theta_2}{2}})-\frac{2q}{4}\cos^2{\frac{\theta_1}{2}}\sin^2{\frac{\theta_2}{2}}\\
&&+\frac{1+q}{4}\sin^2{\frac{\theta_1}{2}}(\cos^2{\frac{\theta_2}{2}}+\sin^2{\frac{\theta_2}{2}})-\frac{2q}{4}\sin^2{\frac{\theta_1}{2}}\cos^2{\frac{\theta_2}{2}}\\
&&+\frac{q}{2}\sin{\frac{\theta_1}{2}}\cos{\frac{\theta_1}{2}}\sin{\frac{\theta_2}{2}}\cos{\frac{\theta_2}{2}}[e^{-i(r_1+r_2)}+e^{i(r_1+r_2)}]\\
&=&\frac{1+q}{4}-\frac{q}{2}[\cos^2{\frac{\theta_1}{2}}\sin^2{\frac{\theta_2}{2}}\\
&&+\sin^2{\frac{\theta_1}{2}}\cos^2{\frac{\theta_2}{2}}+2\cos(s_1+s_2)\sin{\frac{\theta_1}{2}}\cos{\frac{\theta_1}{2}}\sin{\frac{\theta_2}{2}}\cos{\frac{\theta_2}{2}}]\nonumber\\
&\geq&\frac{1+q}{4}-\frac{q}{2}[\cos{\frac{\theta_1}{2}}\sin{\frac{\theta_2}{2}}-\sin{\frac{\theta_1}{2}}\cos{\frac{\theta_2}{2}}]^2\\
&\geq&\frac{1+q}{4}.
\end{eqnarray}

\end{document}